\documentclass[12pt]{iopart}
\usepackage{graphicx}

\begin{document}

\title[Conservation of energy with 5 dimensions]{The effect of a fifth large-scale space-time dimension on
the conservation of energy in a four dimensional Universe}

\author{M B Gerrard and T J Sumner}

\address{Department of Physics, Imperial College London, Prince Consort Road,
London SW7 2BW, UK}
\eads{\mailto{Michael.Gerrard@partnershipsuk.org.uk},
\mailto{t.sumner@imperial.ac.uk}}
\begin{abstract}
The effect of introducing a fifth large-scale space-time dimension
to the equations of orbital dynamics was analysed in an earlier
paper by the authors. The results showed good agreement with the
observed flat rotation curves of galaxies and the Pioneer Anomaly.
This analysis did not require the modification of Newtonian
dynamics, but rather only their restatement in a five dimensional
framework. The same analysis derived an acceleration parameter
$a_r$, which plays an important role in the restated equations of
orbital dynamics, and suggested a value for $a_r$. In this companion
paper, the principle of conservation of energy is restated within
the same five-dimensional framework. The resulting analysis provides
an alternative route to estimating the value of $a_r$, without
reference to the equations of orbital dynamics, and based solely on
key cosmological constants and parameters, including the
gravitational constant, $G$. The same analysis suggests that: (i)
the inverse square law of gravity may itself be due to the
conservation of energy at the boundary between a four-dimensional
universe and a fifth large-scale space-time dimension; and (ii)
there is a limiting case for the Tulley-Fisher relationship linking
the speed of light to the mass of the universe.
\end{abstract}

\pacs{04.50+h}
\maketitle

\section{Introduction}
In an earlier paper [1] we introduced a fifth large-scale space-time
dimension, $r$ to Newton's Second Law, as applied to systems with
angular velocity. The resulting analysis of the orbital motion of
galaxies, which considered only the role of baryonic matter, is
consistent with their observed rotation curves and the Tulley-Fisher
relationship. The dimension $r$, is orthogonal to the three space
dimensions $s(x,y,z)$ and the time dimension, $t$ of a
four-dimensional universe, but does not represent a degree of
freedom of motion in this analysis. For a closed isotropic universe,
$r$ is the radius of curvature of (four-dimensional) space-time and
has a value, $r_u$ remote from gravitating matter that is estimated
to be $\sim7.5\times10^{26}\,$m.  The parameter $a_r$ is derived
from the relationship $a_r = c^2/r$. In the case of $r$ being equal
to $r_u$, $a_r$ has a value of $1.2\times10^{-10}\,\rm{ms}^{-2}$,
which is the same as the MOND parameter $a_0$ derived by Milgrom [2]
from observing the rotation curves of more than eighty galaxies.

Using the same five-dimensional analytical framework, this paper
examines the relationships between $a_r$, the principle of
conservation of energy and gravity. The resulting derivation of
$a_r$ is, therefore, unrelated to orbital dynamics and Newton's
Second Law and instead relies on key cosmological constants, such as
the gravitational constant, $G$ and parameters, such as the mass
density of the universe.

\section{Background Gravitational Acceleration in the Universe}
The large-scale distribution of matter across the universe creates a
background gravitational acceleration, $g_b$ which is isotropic if
matter itself is evenly distributed on this scale. The mutual
attraction of each particle of matter towards all other matter, as
represented by $g_b$, is similar in concept to a three dimensional
``surface tension'' stretching across the universe.

If space is assumed to be flat and open and matter is assumed to be
evenly distributed on this large scale, with (baryonic) mass density
$\rho$, then the background gravitational acceleration, $g_b$, can
be derived as follows:
\begin{equation}
g_b=\pi G \rho H_H
\end{equation}
where $G$ is the gravitational constant
($6.67\times10^{-11}\,\rm{m}^3\rm{Kg}^{-1}\rm{s}^{-2}$), $\rho$ for
baryonic matter has a currently estimated value $\rho_u = 3\times
10^{-28}\,\rm{Kg\,m}^{-3}$ and $H_H$ is the Hubble Horizon given by
$H_H = c/H$ with H being Hubble's Constant
($71\,\rm{Km\,s}^{-1}\rm{Mpc}^{-1}$). Substitution in equation (1)
gives a current value for $g_b$ of $8.2\times10^{-12}\,\rm{ms}^{-2}$
which is noted to be two orders of magnitude less than the value of
$a_0$.

The accuracy of equation (1) depends on three potential sources of
uncertainty, namely: the value of $\rho$, the method of calculation
of the volume of the universe within the Hubble Horizon and the
value of $H$ itself. These will be discussed later.

\section{Background Radius of Curvature of the Universe}
In section 3.2 of the earlier paper [1] an expression was derived
for the locus of points $r(x)$ adjacent to a gravitating mass, $M$
which defined the balance condition between gravitational
acceleration $g_x$ and the acceleration $a_r$ acting everywhere in
the universe in the direction of $r$.
\begin{equation}
    r\left(x\right) = r_u \left(1 - \frac{GM}{c^2x}\right)
\end{equation}
where $r_u$ is the radius of curvature of four-dimensional
space-time {\it remote} from gravitating matter $M$ and $x$ is the
distance away from $M$ as shown in figure 1.

\begin{figure}[t]
\begin{center}
\includegraphics[height=3.5in,angle=270,clip=]{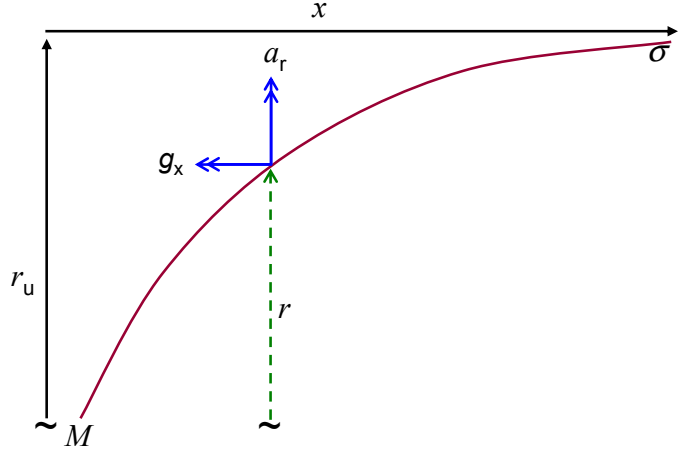}
\caption{Locus of points $r(x)$ at which there is balance between
the two accelerations $g_x$ and $a_r$.} \label{fig1}
\end{center}
\end{figure}

The effect which matter has on the local radius of curvature of
space-time, $r$ is cumulative and can be found by the superposition
($\Delta r/r_u = \Sigma\,\Delta r_i/r_u$, where $\Delta r_i =
\left(r_u - r_i\right)$) from all individual masses, $M_i$. Applying
equation (2) to all baryonic matter contained within the Hubble
Horizon (again assumed to be evenly distributed across space with
density $\rho$ and lying within a spherical volume defined by
$4/3\left(\pi s^3\right)$ where $s$ here is $H_H$) it is possible to
calculate an overall background value of $r(x)$. This value will
inevitably be somewhat less than $r_u$ given that no point is, in
practice, completely remote from all matter. This background value
of $r$ is referred to as $r_b$ and is derived by integrating the
contributions from matter lying within concentric spherical shells
of space to give:
\begin{equation}
    r_b = r_u \left(1 - \frac{2\pi G\rho H_H^2}{c^2}\right)
\end{equation}

Substituting for known parameters and constants in equation (3),
including the current value of the mass density $\rho_u$, gives a
value for $r_b$ equal to $0.98\times r_u$. Substituting either value
for $r$ into the key relationship $a_r = c^2/r$ gives the same value
for $a_r$ to within one decimal place, namely
$1.2\times10^{-10}\,\rm{ms}^{-2}$.

The average mass density of the universe, $\rho$, decreases over
time in an expanding universe. For a Euclidean (although expanding)
universe, the volume of space within the Hubble Horizon is given by
$\left(4/3\right)\pi H_H^3 \simeq \left(4/3\right)\pi (ct)^3$. Given
that (to a first order) the total mass lying within the Hubble
Horizon is constant, it follows that we can derive an expression for
the average mass density $\rho(t)$ of the universe at any time $t$,
in terms of the average mass density observed for the current era
$\rho_u$ (i.e. $\sim 3\times10^{-28}\,\rm{Kg\,m}^{-3}$) and the
current estimated age of the universe $t_u$ (i.e. $13.7\,$Bn years).

\begin{equation}
\rho \simeq \frac {\rho_ut_u^3}{t^3}
\end{equation}

Given that this equation is derived (in part) from the approximation
$H_H \simeq ct$, it is assumed only to be applicable in the current
analysis for perturbations of time about the current era.

Substituting for $\rho$ from equation (4) into equation (3) provides
an expression for the local time-dependency of the background radius
of curvature of space-time $r_b$ in equation (5), which is similarly
limited in its range of extrapolation.

\begin{equation}
r_b = r_u \left(1-\frac {2\pi G\rho_ut_u^3}{t}\right)
\end{equation}

\section{Conservation of Energy}
In section 3.1 in the earlier paper [1] $a_r$ was described as a
universal acceleration of expansion acting at all points in space in
the direction of $r$. To maintain conservation of energy within
four-dimensional space-time, it follows that for any mass $m$ at a
point in space P there must be an acceleration equal and opposite to
$a_r$ which prevents energy being transferred from within the
four-dimensional universe to the fifth dimension $r$, as shown in
figure~2. Accordingly, this principle may be written as:
\begin{equation}
a_r + \frac{d^2r_b}{dt^2} = 0
\end{equation}

The second term of this equation ($\ddot{r}_b$) is identified as the
acceleration acting on a mass in the direction of the dimension $r$
(decreasing) by virtue of the expansion of the universe in the
dimension $r$ which causes $r_b$ the background value of $r$ to
increase over time (but at a decelerating rate $-$ see equation
(5)). In other words, given that the universal acceleration $a_r$ is
acting everywhere along the boundary between the four-dimensional
space-time and the fifth dimension $r$, energy can only be conserved
(within four dimensional space-time) if the background radius of
curvature of space-time $r_b$ varies in time so as to satisfy
equation (6). This conservation of energy at the boundary between
the four dimensional universe and the fifth dimension $r$ is, of
course, the reason why the dimension $r$ is not itself directly
observable.  As referred above, for the current era $a_r$ is
$1.2\times10^{-10}\,\rm{ms}^{-2}$.

Assuming only $r_b$ varies with time equation (5) gives:
\begin{equation}
\frac{d^2r_b}{dt^2} = - 4\pi\rho Gr_u
\end{equation}
Substituting values for known parameters and constants on the
right-hand side (including the current mass density of the universe,
$\rho_u$, provides the result:  $\ddot{r}_b =
-1.9\times10^{-10}\,\rm{ms}^{-2}$. Given the approximations used to
derive equation (7), this value for $\ddot{r}_b$ appears to be in
reasonably good agreement with the value expected from equation (6),
namely: $-1.2\times10^{-10}\,\rm{ms}^{-2}$).

The substitution for $r_u$ in equation (7) using the relationship
$a_r = c^2/r$ (section (1) above), but with the identification of
$a_r = a_o$ for $r=r_u$ for the current era, and the combining of
equations (6) and (7) allows an expression for $a_o$ as:
\begin{equation}
a_o = \left(4\pi \rho_u Gc^2\right)^{1/2}
\end{equation}
which has the value of $1.5\times 10^{-10}\,$ms$^{-2}$ for current
era.

The level of agreement between $a_r$ and $\ddot{r}_b$, calculated
from equation (7), can only be properly assessed by considering the
uncertainty in the three key components to equation (7): the value
of the Hubble Horizon, the average mass density of the universe and
the estimated volume of the universe. Consistency between
$\ddot{r}_b$ from equation (7) and equation (6)) lies within the
uncertainty ranges of $\pm 12\%$ in each of these three components.
However, the principal source of uncertainty in $\ddot{r}_b$ is
expected to be the method used to calculate the volume of the
universe lying within the Hubble Horizon.

\begin{figure}[t]
\begin{center}
\includegraphics[height=3.5in,angle=270,clip=]{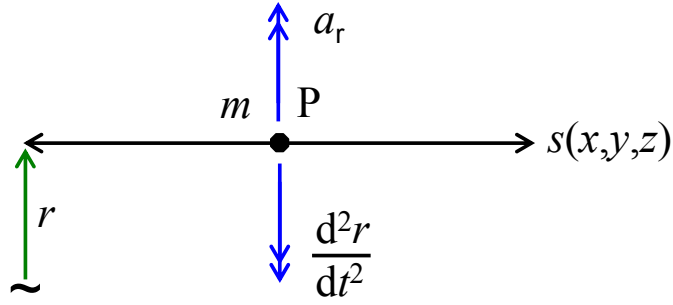}
\caption{Conservation of energy requires the two accelerations $a_r$
and $\ddot{r}_b$ to be equal and opposite.} \label{fig2}
\end{center}
\end{figure}

The form of universe that underpins the derivation of $a_r$ is
closed (i.e. curved) and isotropic {Section 3.2 in [1]} and yet, so
far in this paper, we have used the Euclidean derivation of a three
dimension spherical volume $4/3\left(\pi s^3\right)$, where $s$ is
the radius of the volume - i.e. a derivation appropriate to a flat
and open universe. A closed isotropic three dimensional space is the
``surface area'' of a 4-dimensional hyper-sphere, the 3-dimensional
volume of which is given not by $4/3\left(\pi s^3\right)$ but by the
expression $2\pi^2 R^3$, where $R$ is the radius of curvature of the
hyper-sphere. The relevant feature of this 3-dimensional ``surface
area'' is that at increasing distances $s$ from a point P, the
volumes of concentric spherical shells of space centred on P become
progressively smaller than those derived from the (Euclidean)
expression $4\pi s^2ds$.

Accordingly, failure to take account of this effect will have led to
an over-estimation of the volume of the universe lying within the
Hubble Horizon and so to an over-estimation of $\ddot{r}_b$ in
equation (7). The value of $g_b$ in equation (1) will have,
likewise, been overestimated for this reason.

There are two important aspects of the application of $\rho$ in the
calculation of $\ddot{r}_b$ and $g_b$ that also need to be
highlighted: the first in relation to a closed universe; and the
second in relation to an expanding universe.

\subsection{A closed universe}
The application of a single average value for $\rho$ to a closed
universe, defined by the 3-dimensional ``surface area'' of a
4-dimensional hyper-sphere, means that the contributions of matter
lying within ever more distant volumes of space\footnote{i.e. the
volume of concentric shells of space centred on point P and lying at
distance $s$ from P depart increasingly from $4\pi s^2ds$ as $s$
increases} to the measured values of $\ddot{r}_b$ and $g_b$, will
ultimately diminish with distance. Consequently, inaccuracies in the
value of $H$ and, thereby, the Hubble Horizon should not be primary
sources of error in $\ddot{r}_b$ and $g_b$. Moreover, recent
observations that indicate lower values for $H$ at the furthest
distances should not, for the same reason, undermine the validity of
using a single value for $H$ in the derivation of equations (1) or
(7).

\subsection{An expanding universe}
The nature of expansion of the universe (whether open or closed)
that is assumed here, is one in which mass density is determined by
a fixed amount of matter lying within the Hubble Horizon assumed to
be receding at the speed of light.  To a first order it is not
affected by mass flows across either the Hubble Horizon, or across
regions of space lying within the Hubble Horizon, nor by the
inter-change between matter and energy. Accordingly, a profile of
steadily increasing mass densities at further distances from a point
$P$, due to these further distances being observations of the
universe's past, should not affect the determination of $\ddot{r}_b$
and $g_b$, to the extent that greater mass densities (in the past)
are off-set by reductions in the volume of space (in the past).

If the same adjustment for space being closed as would be needed to
bring to $\ddot{r}_b$ into equality with $a_r$ in equation (6) is
also applied to the derivation of $g_b$ in equation (1), $g_b$
reduces by circa 25\% to $6.0\times10^{-12}\,\rm{ms}^{-2}$. Having
made the same correction for volume, the relationship between the
background value for the radius of curvature of space-time $r_b$ and
$r_u$ also remain unchanged (to one decimal place), namely $r_b =
0.98\times r_u$. Hence, the corrected calculation of the volume of
space lying within the Hubble Horizon does not affect the calculated
value for $a_r$, which remains $1.2\times10^{-10}\,\rm{ms}^{-2}$
(i.e. the same as $a_0$).

Hence, if account is taken of a closed and isotropic nature of space
in applying the current value for the mass density of the universe
$\rho_u$, then the principle of conservation of energy appears to
offer an alternative approach to the valuation of $a_r$ and,
moreover, an approach that is based on key cosmological parameters
and the gravitational constant $G$ and that is independent of
orbital dynamics and Newton's Second Law used in the earlier paper.

\section{Discussion}

A number of simplifying assumptions have been made in this paper.
These include assumptions about the Hubble Horizon, the mass density
of the universe and the calculation of volumes of space over large
distances. Nonetheless, the value for $a_r$ derived from the
principle of conservation of energy is in good agreement with that
expected from MOND observations [2] and from the derivation based on
the Hubble Constant [1].

The relative dominance of proximate matter over very distant matter
in the determination of the background universal gravitational
acceleration $g_b$ and in the background value for the radius of
curvature of space-time $r_b$ (assuming matter is evenly distributed
on a very-large scale and the universe is closed), should make the
calculations used in this paper reasonably robust to inaccuracies in
the estimation of the Hubble Horizon and of volumes of space at
greater distances.

The time dependencies of $r_b$ evident in equation (5) (i.e.
increasing with age of the universe) and of $|\ddot{r}_b|$ evident
in equation (7) (i.e. decreasing with age of the universe) imply
that we should modify the central equation for $a_r$ proposed in the
earlier paper and write it as:
\begin{equation}
    a_r = \frac{c^2} {r_b}
\end{equation}

For a value of $r_b = 0.98\times r_u$, the value of $a_r$ derived
from equation (9) remains the same as $a_0$ (the MOND parameter) to
one decimal place (i.e. $1.2\times10^{-10}\,\rm{ms}^{-2}$) for the
current era. The substitution of $r_b$ for $r_u$ in the equation for
$a_r$ and the principle of conservation of energy (i.e. equation
(6)) are consistent with higher values for $\rho$, $|\ddot{r}_b|$
and $a_r$ in earlier ages of the universe. The observations of
rotation curves of galaxies which support the MOND parameter $a_0$
proposed by Milgrom have, so far, mostly covered galaxies out as far
as $\sim100\,$Mpc from earth. To one decimal place, there is no
change to $r_b$ from equation (5) over these distances and so no
corresponding departure from the MOND value for $a_0$ would be
expected from equation (9).

The analysis in sections 3 and 4 can, of course, be reversed and the
principle of conservation of energy as expressed by equation (6) can
be used as the starting point to derive the underlying relationship
between matter and the radius of curvature of 4-dimensional
space-time in an expanding universe, namely equation (2). If this
approach is adopted, then the inverse square law of gravity (which
is a derivative of equation (2)\footnote{For the relationship
between equation (2) and the inverse square law of gravity, see
section 3.2 [1]}) may be inferred as a consequence of the
conservation of energy at the boundary between a (closed) expanding
four-dimensional universe and a fifth large-scale dimension of
space-time. Accordingly, a description of gravity based upon this
principle of conservation of energy would appear to offer a
derivation based on thermodynamics for the key dimensionless term of
General Relativity ($GM/c^2x$). Furthermore, equations (7) and (9)
may be substituted in equation (6) to provide an expression for the
gravitational constant (G), of the following form:
\begin{equation}
    G = \frac{kc^2r_u} {M_{universe}}
\end{equation}

where $M_{universe}$ is the mass of the universe and $k$ is a
dimensionless constant determined by the correct approach to
calculating the volume of the universe. This equation suggests a
link between $G$ and the key fifth dimensional parameter $r_u$,
which is identified in this and the earlier paper as the radius of
curvature of space-time remote from gravitating matter; albeit with
the same limitations as equation (7) from which it is derived. All
the terms on the right-hand side of equation (10) are, as expected,
constant.

Finally, equation (10) can itself be restated in terms of the
parameter $a_o$ rather than $r_u$ by substituting the expression
$a_o = c^2/r_u$:
\begin{equation}
c^4 = a_oGM_{universe}k^{-1}
\end{equation}
which is of the form of the Tulley-Fisher relationship (see equation
(25) in [1]).   The equation suggests a limiting case for this
relationship and, moreover, one which is derived from the principle
of conservation of energy at a universal level and without reference
to the orbital dynamics of individual galaxies or the universe as a
whole.

\end{document}